\documentclass[tikz,conference]{IEEEtran}
\IEEEoverridecommandlockouts
\usepackage{bm}
\usepackage{cite}
\usepackage{amsmath,amssymb,amsfonts}
\usepackage{algorithmic}
\usepackage[linesnumbered,ruled,vlined]{algorithm2e}
\usepackage{graphicx}
\usepackage{textcomp}
\usepackage{xcolor}
\usepackage{subfigure}
\def\BibTeX{{\rm B\kern-.05em{\sc i\kern-.025em b}\kern-.08em
    T\kern-.1667em\lower.7ex\hbox{E}\kern-.125emX}}
    
\usepackage{amsthm}

\usepackage{tikz}
\usetikzlibrary{shapes.geometric, arrows, positioning,calc}

\tikzstyle{startstop} = [rectangle, rounded corners, minimum width=3cm, minimum height=1cm,text centered, draw=black, fill=red!30]
\tikzstyle{io} = [trapezium, trapezium left angle=70, trapezium right angle=110, minimum width=3cm, minimum height=1cm, text centered, draw=black, fill=blue!30]
\tikzstyle{process} = [diamond, minimum width=3cm, minimum height=1cm, text centered, draw=black, fill=orange!30]
\tikzstyle{decision} = [diamond, aspect=2, text centered, draw=black, fill=green!30]
\tikzstyle{arrow} = [thick,->,>=stealth]

\allowdisplaybreaks

\title{
Towards Accurate Ego-lane Identification with Early Time Series Classification
}
\author{Yuchuan Jin, Theodor Stenhammar, David Bejmer, Axel Beauvisage, Yuxuan Xia, and Junsheng Fu
\\
Zenseact, Gothenburg, Sweden
\\
Email: firstname.lastname@zenseact.com
}

\begin{document}

\maketitle
\thispagestyle{empty}
\pagestyle{empty}

\begin{abstract}
Accurate and timely determination of a vehicle's current lane within a map is a critical task in autonomous driving systems. This paper utilizes an Early Time Series Classification (ETSC) method to achieve precise and rapid ego-lane identification in real-world driving data. The method begins by assessing the similarities between map and lane markings perceived by the vehicle's camera using \textit{measurement model quality} metrics. These metrics are then fed into a selected ETSC method, comprising a probabilistic classifier and a tailored trigger function, optimized via multi-objective optimization to strike a balance between early prediction and accuracy. Our solution has been evaluated on a comprehensive dataset consisting of 114 hours of real-world traffic data, collected across 5 different countries by our test vehicles. Results show that by leveraging road lane-marking geometry and lane-marking type derived solely from a camera, our solution achieves an impressive accuracy of 99.6\%, with an average prediction time of only 0.84 seconds.

\end{abstract}
\section{Introduction}
Localization is an critical module for autonomous vehicles to navigate safely and efficiently in real-world environments. Traditionally, vehicle localization has relied on the fusion of sensor data of diverse types, such as visual, inertial, or Global Navigation Satellite System (GNSS) inputs, to yield pose estimations. 
Over the past decades, a range of algorithms, including e.g., INS/GNSS navigation \cite{Almagbile2011,chiang2020performance,he2023research}, Visual Odometry (VO) \cite{Nister2004a,zhan2020visual}, Simultaneous Localization and Mapping (SLAM) \cite{Ila2016,zheng2023simultaneous}, and Visual-Inertial Navigation Systems (VINS) \cite{Huang2019}, have been developed to address the challenges in localization.
Nevertheless, these techniques may encounter challenges when applied to large datasets. For instance, Visual Odometry (VO) and SLAM algorithms often exhibit drift over time in the absence of loop closure strategies \cite{Huang2019}. Additionally, GNSS can have substantial biases in urban environments, leading to significant degradation in localization accuracy \cite{Cai2018IntegrationLocalization}.
In recent years, the adoption of High Definition (HD) maps has increased in popularity, primarily due to their ability to offer support for precise localization \cite{Liu2020HighAnalysis, Jo2017UsingUncertainty, Li2017Lane-levelMaps, Cai2018IntegrationLocalization}. An HD map in our context is a semantic map that contains lane-level information such as lane markings, barriers, and traffic signs with positioning and types information \cite{caesar2020nuscenes, wilson2023argoverse}. 

Identifying the correct ego-lane within an HD map by leveraging vehicle onboard sensors is a crucial task during the initialization phase of HD map-based localization. In our context, the initialization phase spans from the moment the automated driving system receives sensor measurements and map data to the point at which the system selects an ego-lane for the ego vehicle.
Initialization (or re-initialization) can occur for various practical reasons and at different locations, such as when sensor measurements are lost for an extended period, navigating through areas with low-quality mapping (e.g., road work), or traversing unmapped regions. 
Ego-lane identification poses particular challenges in areas with numerous lanes or even multiple road layers as GNSS could fail to provide sufficient accuracy to identify the specific lane the vehicle occupies\cite{Laconte2022AScenarios}.

Road lane markings are ubiquitous features on most roads, and HD maps meticulously capture both their geometry and types. Additionally, lane markings can be quite easily detected by a vehicle's camera. By utilizing GNSS measurements as the rough estimate of the vehicle's position and leveraging the perceived geometry and types of lane-markings captured by the vehicle's camera, it is possible to identify the current ego-lane within an HD map. This paper focuses specifically on ego-lane identification during the initialization phase by using the road lane-marking geometry and types.

\begin{figure}
    \centering
    \includegraphics[width=\columnwidth]{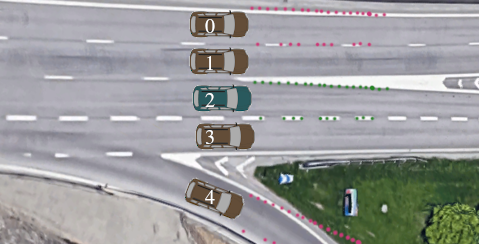}
    \caption{An illustration depicts perceived lane markings from various hypotheses' viewpoints. The red dots represent perceived lane markers from Hypotheses 0 and 4, which do not align well with the map. In contrast, the green dots depict lane markers perceived from Hypothesis 2 (the correct one), showing a more accurate alignment with the map.
    }
    \label{fig:hypothesis_comparison}
\end{figure}

Several techniques based on multi-hypothesis tracking have been proposed to address the ego-lane identification problem \cite{Reid1979AnTargets,Jensfelt2001ActiveTracking,Abdallah2011ALocalization}. These approaches initially generate hypotheses in various lanes. Subsequently, they jointly track the location of these hypotheses across different lanes by incorporating sensor measurements, aiming to determine which one accurately represents the vehicle's position. Fig. \ref{fig:hypothesis_comparison} illustrates the process of generating multiple hypotheses along different lanes on a highway. The automated driving system then utilizes the camera-perceived road lane-marking geometry and types to identify the correct ego-lane in an HD map. In this scenario, hypothesis 2 is determined to be the correct one. Accomplishing this task presents a significant challenge due to the necessity of identifying the correct hypothesis with a high level of confidence, and this confidence naturally grows with the accumulation of more observations. However, it should also avoid unnecessary delays. Therefore, striking a balance between accuracy and promptness is essential, a task commonly referred to as the Early Time Series Classification (ETSC) problem. 

In this paper, we utilize an effective ETSC solution for ego-lane identification in HD map with lane-marking geometry and lane-marking types. For camera perceived lane-marking geometry and lane-marking types,  we compute the similarity metric called \textit{measurement model quality} (MMQ, refer details to subsection \ref{sec:mht}) by comparing perceived lane-marking's geometry and types against the HD map. These metrics are then inputted into a selected ETSC solution, which comprises a probabilistic classifier and a tailored trigger function. This solution is then further optimized through multi-objective optimization to strike a balance between early prediction and accuracy.

The main contributions of this paper are:
\begin{enumerate}
    \item We conducted an evaluation and adaptation of the ETSC method for the ego-lane identification task. The results indicate that our modified approach shows improvements in terms of promptness and availability.
    \item Our experiments utilized 114 hours of real-world driving data to assess the efficacy of the ETSC solution in the domain of ego-lane identification, confirming its robust performance.
\end{enumerate}

The rest of the paper is organized as follows. In Section \ref{sec:rw}, we summarize existing work in multiple-hypothesis tracking and the ETSC problem. Section \ref{sec:method} outlines the entire methodology we employ to address the multiple-hypothesis ETSC problem. Our experimental procedures and their results are detailed in Section \ref{sec:experiments}, while Section \ref{sec:conclusions} delves into the discussions and conclusions drawn from our findings.

\section{Related Work}
\label{sec:rw}
Multiple-hypothesis tracking is a classic method for tracking multiple targets by propagating multiple data association hypotheses over time \cite{Reid1979AnTargets}. The idea of maintaining multiple hypotheses at a time has been adapted to the global localization problem (also commonly known as the \textit{kidnapped robot} problem) to localize an autonomous agent in its environment without prior knowledge \cite{Jensfelt2001ActiveTracking}. In addition, the use of digital maps can be incorporated into the framework for automotive navigation purposes \cite{Abdallah2011ALocalization,Jong-SunPyo2001DevelopmentTechnique}. 



The multi-hypothesis localization problem can be formulated as an ETSC task since the data is continuously collected by the vehicle and evaluated regularly with the goal to reduce the uncertainties captured by multiple hypotheses as early as possible. Then, similar to ETSC \cite{Santos2017ALearning}, machine learning techniques can be leveraged in solving multi-hypothesis localization. In machine learning, classification of time series data differs from classification of tabular data in the sense that sequential information are being processed rather than individual snapshots \cite{Bagnall2017TheAdvances}. Therefore, it is important to employ the right machine learning method that fits the type of data in our specific application \cite{Ruiz2021TheAdvances}.

A practical issue with building an ETSC model is that the time series data used as model input are of different lengths. This often requires a different approach to classification than offline time series classification. The reason for this is that in ETSC the full time series is not available. Some methods for time series classification, such as the Multi-Layer Perceptron proposed in \cite{Wang2017TimeBaseline}, depend on the length of time series. A method that is independent of the length of the time series can be used any time new data is collected, but a method that is dependent on the length can only be used for the specific time series length that it is constructed for. The application of such length-dependent methods for ETSC requires first defining a set of lengths, at which time series are evaluated, and then constructing a different classifier for each possible length of the time series data. 

To overcome the limitation of length-dependent methods, a two-tiered structure has been proposed in \cite{Mori2019EarlyTechniques,Schafer2020TEASER:Classification}, which first divides a time series into segments of different lengths and then uses a classifier trained to output class probabilities for the different lengths of time series segments. Depending on the prediction certainty, a reject-classifier either chooses to output a prediction or wait for more data. Recently, it has been observed in \cite{achenchabe2023classify} that methods for ETSC problems are sensibly adapted in the Early Open Time Series Classification problem (EOTSC) and the transformation methods of ETSC for EOTSC problem have been illustrated. However, there is still ongoing investigation regarding the trade-off among accuracy, earliness, and availability. In the context of localization tasks, both earliness and availability are crucial factors for timely performance. Therefore, it is imperative to conduct a comprehensive study on EOTSC methods, particularly from the perspectives of earliness and availability.
\section{Method}
\label{sec:method}
In this section, we describe our Swift Multi-objective Early Open Time Series Classification method (SMEOTSC), which can be considered as an extension of the previous methods in \cite{Mori2019EarlyTechniques} and \cite{achenchabe2023classify}, named Multi-objective ETSC (METSC) and Multi-objective EOTSC (MEOTSC), respectively. Our method is tailored to the problem ego-lane identification within an HD map.


\subsection{Hypothesis generation}
\label{sec:hg}
To perform ego-lane identification, we first employ a multiple-hypothesis tracking method. We dynamically generate hypotheses, with a maximum limit of K, placing one hypothesis in each lane on the HD map. This process progresses from the nearest to the farthest lane, determined by the distance between the latest GNSS measurement and the center of each lane.

\subsection{Measurement model quality generation}
\label{sec:mht}
For each hypothesis, we use a Kalman filter to track its in-lane position (\(x, y, yaw\)) by utilizing the detected left and right lane-marking of the ego-lane from the vehicle camera.
At each subsequent time step, and for each hypothesis, we gather various measurements such as lane marking sample points (indicating geometry) and their types (e.g., solid, dashed, doubled). These measurements are then associated with the HD map segments surrounding the latest GNSS measurement. The measurement model defines the relationship between the state of the hypothesis and the lane-markings measurements. 

To assess the likelihood of the different hypotheses and to discern the true one from the others, it is necessary to translate the measurements acquired by the sensors into matching scores. We name such metric \emph{measurement model quality} (MMQ) since they are derived from the measurement model. The MMQs are computed for each sensor measurement, based on the Normalized Innovation Squared (NIS) and the chi-square $\chi^2$ distribution. The NIS value is computed as

\begin{equation}
    \text{NIS} = (z_t-\hat{z}_t)^T S^{-1} (z_t-\hat{z}_t),
\end{equation}
where $z_t$, $\hat{z}_t$ denote the observed measurement and the predicted measurement at time step $t$ respectively; $S$ denotes the innovation covariance matrix computed in the Kalman update; $z_t-\hat{z}_t$ corresponds to the innovation term (also known as residual).

Though each type of measurement possesses its own measurement model and produces its own innovation term, the fact that it is normalized by its associated covariance matrix makes it possible to treat all measurements in the same way. Therefore, the MMQ is a scalar value defined as:
\begin{equation}
\text{MMQ} = - \log \frac{\text{NIS} + penalty}{chi2inv(p, d)}
\end{equation}
where $chi2inv(p, d)$ denotes the inverse cumulative distribution function of $\chi^2$ distribution with $d$ degrees of freedom and evaluated at probability $p$; $penalty$ represents a threshold that depends on the number of outlier measurements.

The motivation for this penalty term is to penalize the case when no association can be made between the measurement and the map element or when their distances are larger than some threshold. As a result, the association cost is normalized based on the number of samples that were used, following a $\chi^2$ distribution, where the degrees of freedom of the $\chi^2$ distribution corresponds to the number of samples.

The MMQs are treated as a matching cost, and they reflect how well the perceived measurement reflects the system expectation from its estimated state. The $\chi^2$ normalization ensures that all MMQs have similar magnitudes, regardless of their number of samples. This way, it is easier to combine model qualities and compare hypotheses. For example, a traffic sign position should produce a similar MMQ value to a lane marking represented by 10 points if their innovation term is equally consistent with their uncertainty (variance). These MMQs values are used as derived features in the subsequent learning algorithms.

\subsection{Learning-based hypothesis inference}
\label{sec:lbhi}
\begin{figure}
\centering
\begin{tikzpicture}[node distance=2cm,scale = 0.6,transform shape]
\node (ts1) [io] {$TS^t_1$};
\node (prob1) [process, below of=ts1,yshift=-1.5cm] {Classifier};
\node (p1) [startstop, below of=prob1,yshift=-1.5cm] {$p_1$};
\node (ts2) [io, right of=ts1, xshift=2cm] {$TS^t_2$};
\node (prob2) [process, below of=ts2,yshift=-1.5cm] {Classifier};
\node (p2) [startstop, below of=prob2,yshift=-1.5cm] {$p_2$};
\node (dots) [right of=ts2, xshift=1cm] {$\cdots$};
\node (dots2) [right of=p2, xshift=1cm] {$\cdots$};
\node (tsn) [io, right of=dots, xshift=1cm] {$TS^t_n$};
\node (probn) [process, below of=tsn,yshift=-1.5cm] {Classifier};
\node (pn) [startstop, below of=probn,yshift=-1.5cm] {$p_n$};
\node (sort) [process, below of=p2, yshift=-1cm] {Sorting};
\node (trigger) [decision, below of=sort, yshift=-1.5cm] {Trigger function};
\node (output) [startstop, below of=trigger, yshift=-1.5cm] {Output of most probable hypothesis};
\draw [arrow] (ts1) -- (prob1);
\draw [arrow] (prob1) -- (p1);
\draw [arrow] (ts2) -- (prob2);
\draw [arrow] (prob2) -- (p2);
\draw [arrow] (tsn) -- (probn);
\draw [arrow] (probn) -- (pn);
\draw [arrow] (p1) -- (sort);
\draw [arrow] (p2) -- (sort);
\draw [arrow] (pn) -- (sort);
\draw [arrow] (sort) -- (trigger);
\draw [arrow] (trigger) -- node[anchor=east] {True} (output);
\draw [arrow] (trigger) -| node[near start, below] {False} ($(tsn.east) + (1,0)$) |- (tsn.east);
\draw [arrow] (trigger) -| node[near start, above] {Wait for more data} ($(tsn.east) + (1,0)$) |- (tsn.east);
\end{tikzpicture}
      \caption{An overview of the ETSC architecture, based on the process described in \cite{Mori2019EarlyTechniques}. $TS^t_n$ is the $n$:th Time Series (TS) at time $t$; $p_n$ is the predicted true-probability given by classifier for hypothesis $n$; Trigger function is in charge of multi-hypotheses selection.}
      \label{fig:model_arc}
\end{figure}
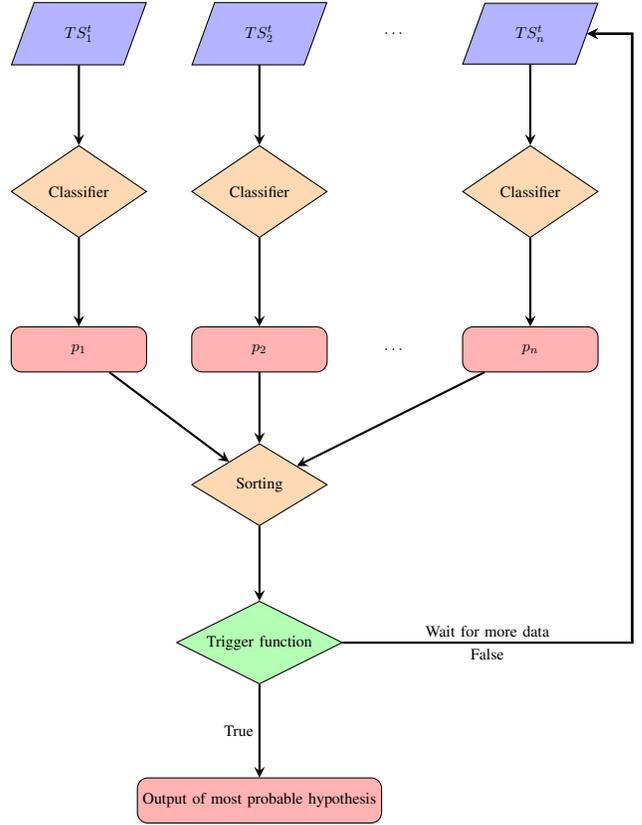

The full architecture of the hypothesis inference solution is shown in Fig. \ref{fig:model_arc}. The input to the model is a set of time series $\bm{X}=(\bm{x_1},\bm{x_2},...,\bm{x_t})$ from multiple hypotheses. At each time step $t$, $\bm{x_t}=(x^1_t,x^2_t,...,x^m_t)$ is represented as a vector consisting of MMQ values discussed in Section \ref{sec:mht}. The final output is the index of the hypothesis with the highest probability of being correct. The solution can be divided into the following two steps.

First, the multivariate time series data is fed to a set of probabilistic classifiers, which outputs the confidence of hypothesis $n$ at a given time $t$. The classifiers are trained with labeled multivariate time series set $\bm{X}$. Second, the certainty of making a selection is determined by comparing the confidence levels of different hypotheses. The hypotheses' confidences are sorted by magnitude and provided to a trigger function $s_{\bm{\gamma}}$, which determines the appropriate action. It decides whether to make a selection at this moment or if the model requires additional data (time steps) to confidently identify the most probable hypothesis indicated by the classifiers.

The trigger function's role is to assess whether the system can confidently generate a prediction at a given time $t$ based on the probabilities of provided hypotheses, or if additional data is needed. As illustrated in Fig.\ref{fig:model_arc}, the trigger function outputs either True or False based on the classifier's result. If the trigger function determines the prediction to be reliable (outputting True), then the hypothesis with the highest probability will be selected. Conversely, if the trigger function decides that the prediction is not yet reliable (outputting False), further data is then needed before making a reliable prediction.

The trigger function proposed in \cite{Mori2019EarlyTechniques} is shown as follow:
\begin{equation}
    s^1_{\bm{\gamma}}(\bm{p},t) =
    \begin{cases} 
        0, & \text{if} \quad \gamma_1p' + \gamma_2(p'-p'') + \gamma_3\frac{t}{T}\leq0 \\
        1, &\text{otherwise}.
        \label{eq:trigger1}
    \end{cases} 
\end{equation}
where $\bm{p}=(p_1,p_2,...,p_n)$ is a vector of probabilities output from classifiers for each hypotheses; $p'$ and $p''$ are, respectively, the highest and the second-highest hypothesis probabilities obtained at time step $t$; $\bm{\gamma}=(\gamma_1, \gamma_2, \gamma_3)$  is a vector of parameters with $\bm{\gamma}_i \in [-1, 1]$; $t$ is current time step; $T$ is the total time steps of time series input.

The approach from \cite{Mori2019EarlyTechniques}, named METSC, is aimed to solve the multi-class classification problem, where a time series is classified after some number of observations of a time series of indeterminate length. However, as discussed in Section \ref{sec:rw}, this method is constrained in the context of EOTSC problems due to the absence of a total time series duration for such problems. Consequently, \cite{achenchabe2023classify} has adapted the trigger function (\ref{eq:trigger1}) specifically for the EOTSC domain:

\begin{align}
    s^2_{\bm{\gamma}}(\bm{p},t) =&
    \begin{cases} 
        0, & \text{if } \gamma_1p' + \gamma_2p'' + \gamma_3\frac{\eta-t}{\eta}\leq0 \\
        1, & \text{otherwise}
        \label{eq:trigger2}
    \end{cases} 
\end{align}
where $\eta$ is predefined total time steps for the time series input.

As can be seen, trigger function (\ref{eq:trigger2}) employs a predetermined total time steps for EOTSC problems and its value decreases over time. However, the question persists: How should we define the total time steps in advance?  Firstly, determining an appropriate total number of time steps for various types of tasks is challenging, resulting in adding another parameter to be learned, which complicates the learning process. Secondly, this approach retains the time penalty element, which might actually impede the algorithm's ability to determine the reliability of predictions at an early stage. To address these challenges and to tailor to our specific ego-lane identification problem, we adopt trigger function (\ref{eq:trigger2}) as follows:
\begin{align}
    s^3_{\bm{\gamma}}(\bm{p}) &=
    \begin{cases} 
        0, & \text{if } \gamma_1p' + \gamma_2(p'-p'') + \gamma_3\leq0 \\
        1, & \text{otherwise}
        \label{eq:trigger3}
    \end{cases} \\
    s^4_{\bm{\gamma'}}(\bm{p}) &=
    \begin{cases} 
        0, & \text{if } \gamma_1p' + \gamma_2(p'-p'')\leq0 \\
        1, & \text{otherwise}
        \label{eq:trigger4}
    \end{cases} 
\end{align}
where $\bm{\gamma'}=(\gamma_1, \gamma_2)$ is a vector of parameters with $\bm{\gamma}_i \in [-1, 1]$

In trigger function (\ref{eq:trigger3}), $\gamma_3$ is kept to ensure consistency in parameter numbers with (\ref{eq:trigger1}) and (\ref{eq:trigger2}) for comparative purpose. Additionally, trigger function (\ref{eq:trigger4}) also removes parameter $\gamma_3$ to let algorithm give a prediction as early as possible when the classifier results are reliable enough. For the lane-level localization application, it is desired that a prediction is made once the information state is good enough to determine the ego-lane index with certitude. The solution should therefore be reactive to fluctuations in the information state and output as quickly as possible once a selection is feasible and reliable. However, it should exercise restraint in making a selection in instances where different hypotheses are indistinguishable. Detailed analysis related to trigger function (\ref{eq:trigger3}) and (\ref{eq:trigger4}) will be presented in Section \ref{sec:ra}.


\subsection{Multi-objective optimization}
\label{sec:moo}

Three metrics can be utilized to evaluate the performance of the method: availability, earliness, and accuracy. Their definitions are as follows:

\newtheorem{definition}{Definition}
\begin{definition}
    \textbf{Availability} is defined as the fraction of sequences for which a prediction was made within the observation period. In our experiments, we establish a maximum allowable time for generating a prediction. If the system exceeds this time limit without producing a prediction, we consider it a failure in our assessment. Then, the availability can be calculated by dividing the number of successful predictions by the total number of sequences.
\end{definition}
\begin{definition}
    \textbf{Earliness} is defined as the average number of observations needed before a prediction was made, normalized by second.
\end{definition}
\begin{definition}
    \textbf{Accuracy} is defined as the fraction of correct predictions for all sequences where a prediction is available.
\end{definition}

To obtain a robust estimation of the parameter vector $\bm{\gamma}$, a multi-objective function targeting availability and accuracy is optimized in a similar manner as in \cite{Mori2019EarlyTechniques}. The motivation behind optimizing on availability instead of earliness is that the availability of a prediction is more important in automated driving scenarios. Consider a scenario where a driver employs an Advanced Driver Assistance System (ADAS). Typically, the initialization phase occurs automatically when the journey begins. It's common for drivers to cover some distance before utilizing the ADAS features, thus surpassing the initialization phase. Indeed, in cases where no prediction can be made with high confidence, the model should not be forced to predict based on an earliness factor.

The accuracy cost is defined as the prediction error, which is the percentage of incorrectly predicted time series. The earliness cost is defined as the ratio of the time step at which the prediction is made to the total time steps of the time series. The availability cost is defined as the non-prediction rate, indicating the percentage of time series not predicted within the total length of the time series. Their expressions are given as follows:

\begin{align}
    C_{ac}(\bm{X},s_{\bm{\gamma}}) &= \frac{1}{|\bm{X}|}\sum_{\bm{x}\in \bm{X}}{\iota(\hat{y_x}\neq y_x)},
    \label{eq:acc} \\
    C_{ea}(\bm{X},s_{\bm{\gamma}}) &= \frac{1}{|\bm{X}|}\sum_{\bm{x}\in \bm{X}}\frac{t^*_x}{L_x}
    \label{eq:ear_cost}, \\
    C_{av}(\bm{X},s_{\bm{\gamma}}) &= \frac{1}{|\bm{X}|}\sum_{\bm{x}\in \bm{X}}\iota(\nexists{\hat{y_x}}), \label{eq:av}
\end{align}
where $\iota()$ is a boolean function, taking a value of 1 if the condition is true and 0 otherwise; $y_x$ and $\hat{y_x}$ are the ground truth value and prediction value, respectively; $t^*_x$ is the time step when prediction is made for time series $\bm{x}$; and $L_x$ is the total number of time step for time series $\bm{X}$.

The explained cost function regarding availability is similar to the earliness cost function by \cite{Mori2019EarlyTechniques}. The availability cost can be seen as a discrete earliness cost, which only outputs a positive value if no prediction has been made during the sequence. The earliness cost is more continuous in the sense that it returns a growing value from the first time step.

The optimization problem is defined as follows:
\begin{equation}
    \min_{\boldsymbol{\gamma}}(C_{av}(\bm{X},s_{\bm{\gamma}}),C_{ac}(\bm{X},s_{\bm{\gamma}}))
    \label{eq:obj_2}
\end{equation}
The objective of this multi-objective optimization problem is to minimize both the availability cost and the accuracy cost. Following the approach outlined in \cite{Mori2019EarlyTechniques}, we apply meta-heuristic algorithms to address this optimization problem due to its complexity involving multiple objectives.
\section{Experiments}
\label{sec:experiments}
In this section, we present and analysis the experimental results on real-world data.
\subsection{Data}
We gathered 114 hours of experimental data from across 5 countries: Germany, France, Belgium, Sweden, and the United States. The test vehicle has the same setup as the one in our Zenseact Open Dataset \cite{alibeigi2023zenseact}. 
Table \ref{tab:country_statistic} shows the statistics regarding the number of sequences and driving hours for each country. This dataset encompasses diverse measurements, but our experiments have mainly used GNSS and front-looking camera data. The GNSS is employed solely to provide a rough location estimate of the vehicle, facilitating the retrieval of surrounding HD map data. Subsequently, the ego-lane identification only relies on camera data and the retrieved HD map data. The driving scenarios recorded were primarily situated in multi-lane highway settings because the HD map coverage is limited in Highway in our dataset. Visual data, if discernible, facilitated the identification of lane-markings pertinent to both the ego-lane and its adjacent lanes. The dataset further delineated the geometrical characteristics of these lanes, and the quality of the measurement models was evaluated as explained in Section \ref{sec:mht}.

The inputs for ETSC comprise 8 measurement model qualities. 4 of them focus on the geometric properties of both the ego-lane and its adjacent lanes, while the remaining 4 characterize the types of lane markings within the ego-lane. All these measurements are consistently computed at a frequency of 40Hz:
\begin{itemize}
        \item ego-lane left lane-marking geometry
        \item ego-lane right lane-marking geometry
        \item adjacent-lane left lane-marking geometry
        \item adjacent-lane right lane-marking geometry
        \item ego-lane left lane-marking type
        \item ego-lane right lane-marking  type
        \item adjacent-lane left lane-marking type
        \item adjacent-lane right lane-marking  type
\end{itemize}
For any given time step or attribute where data is absent, the measurements are considered missing. Such gaps in data might arise from inconsistencies in the map, or perception discrepancies related to lane markers.

\begin{figure}
    \centering
        \includegraphics[width=0.8\columnwidth]{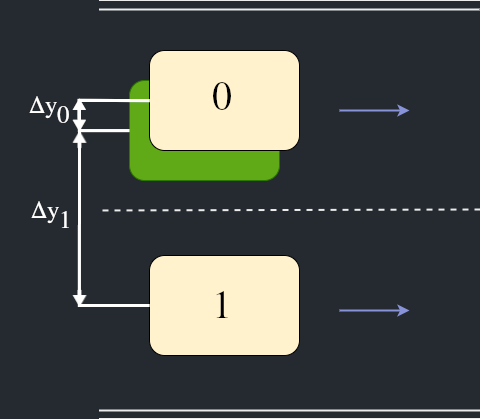}
        \caption{Visualisation of the annotation scheme with the green box being the reference (ground truth). $\Delta y$ is the difference between the center of ground truth and hypothesis. Hypothesis 0 is labeled as true, and hypothesis 1 is labeled as false.}
        \label{fig:gt}
\end{figure}

To train a supervised machine learning model, the availability of ground truth labels is crucial. These ground truth labels are derived from a reference model, which is generated offline by utilizing data acquired from high-precision external sensors during the data collection phase. The sensor setup is same as \cite{alibeigi2023zenseact}, which comprises a set of GNSS/INS positions collected by an OxTS RT3000 unit and this ground truth pose is further refined by aligning Lidar scans with HD map with an ICP \cite{ICP}. The Velodyne sensors are a top-mounted VLS-128 Lidar and two side-mounted VLP-16 Lidars. The experiments are conducted on a laptop with an Intel(R) i7-1365U processor (1.80 GHz), 10 cores, and 32 GB of RAM.

Fig. \ref{fig:gt} illustrates the annotation scheme. The time series are categorized as either \textit{true} or \textit{false} based on their lateral proximity to the reference pose, as established by the reference system. A hypothesis is designated as \textit{true} only if it is the singular hypothesis closest to the reference pose, with all other hypotheses consequently labeled as \textit{false}. A complete time series is deemed valid if every time step for a particular hypothesis is labeled as \textit{true}. Consequently, sequences exhibiting ambiguity in their ground truth are excluded. Such ambiguities may arise from discrepancies in the reference pose or misalignment in the map.

\begin{table}[t]
    \centering
    \caption{Number of sequences and driving hours collected in 5 countries}
    \begin{tabular}{c|c|c}
        \hline
            Country & Number of Sequences & Driving Time (Hours) \\
        \hline
            Germany & 423 & 17.4 \\
            France & 656 & 26.6 \\
            Belgium & 328 & 13.4 \\
            Sweden & 768 & 32.2 \\
            United States & 581 & 24.1 \\
    \end{tabular}
    \label{tab:country_statistic}
\end{table}

\begin{figure}
    \centering
            \includegraphics[width=\columnwidth]{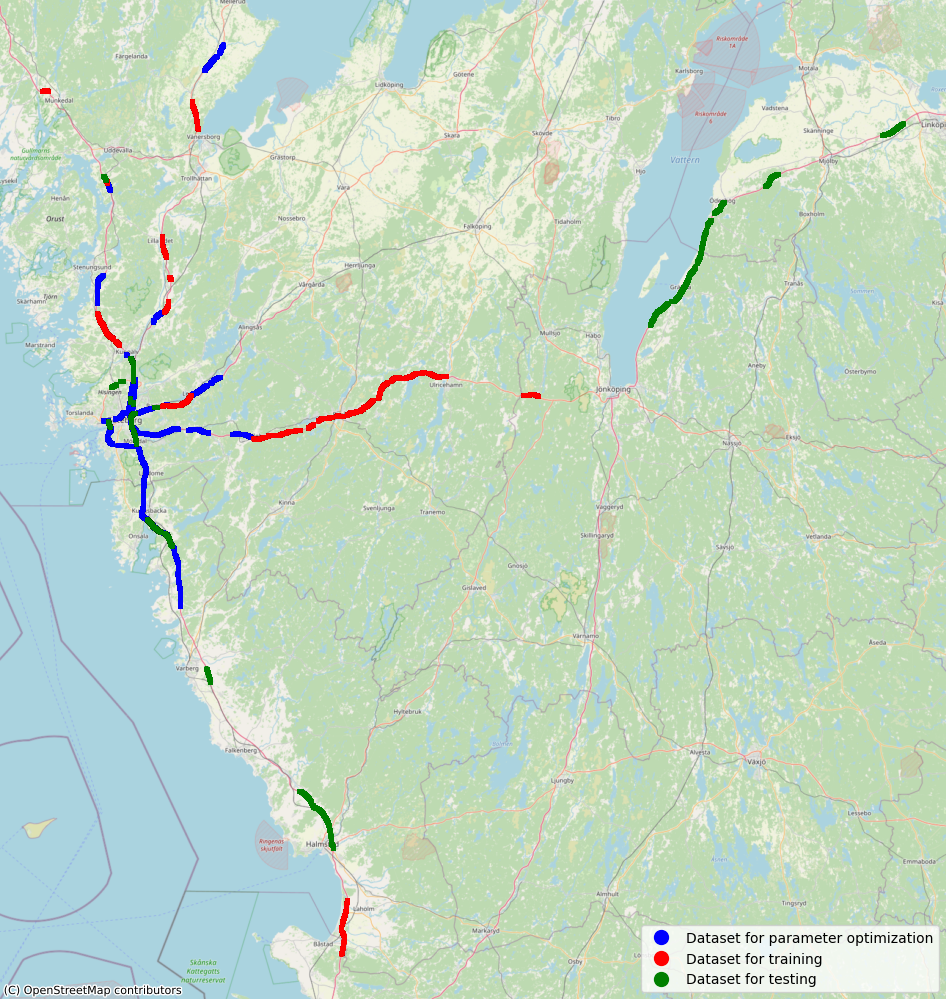}
            \caption{Visualization of the 32-hour driving logs collected around Gothenburg, Sweden. The dataset was divided into 3 groups, red, blue, and green, which have been used for training, parameter optimization, and testing respectively. The map is drawed with openstreet basemap \cite{OpenStreetMap}.}
            \label{fig:datasets}
\end{figure}

\subsection{Experiments Setup}
As mentioned in Section \ref{sec:lbhi} and \ref{sec:moo}, a set of classifiers need to be trained for estimating the probability of different hypotheses at different time and a meta-heuristic optimization algorithm is needed to solve the multi-objective optimization problem. Analyzing the performance of different classifiers and meta-heuristic optimization algorithm to solve this problem is beyond the scope of this paper. In order to perform our experiments, we have chosen gradient boosting algorithm \cite{achenchabe2023classify} as classifier training algorithm and NSGA II optimization algorithm \cite{Mori2019EarlyTechniques} as multi-objective optimization algorithm. 

Gradient Boosting is a powerful boosting algorithm in machine learning used for classification, which combines several weak learners into strong learners. The algorithm has been implemented in Python using the \emph{sklearn} package and the default parameters are used. NSGA II is a fast and elitist genetic algorithm which is established in multi-objective optimization scenarios. The algorithm has been implemented in Python using the \emph{pymoo} package. Except for a parameter change in population size to 8 and termination step of 16, the default parameters are used. The mentioned adjustments to the default parameters have been made to make the algorithm run faster.

The dataset was divided into three subsets for training the classifier, performing the multi-objective optimization, and evaluating the results. Feature extraction for providing the probability estimates was performed by averaging the features over a set of windows with different sizes. There are 2756 sequences and around 114 hours of driving in total. Each sequence has a maximum duration of 30 seconds and contains up to 6 hypotheses. These hypotheses will be generated for each lane on the HD map (up to 6), advancing from the nearest to the farthest lane, determined by the distance between the previous GNSS measurement and the lane center of each lane. Not all hypotheses are present for the full 30 seconds as they may have been deactivated early. An early deactivation of an hypothesis can be due to the hypothesis running out of the map, which happens when there is no map available for the road. The evaluation was performed using the earliness, accuracy, availability metrics as described in Section \ref{sec:moo}. The other performance evaluation metric is hypervolume \cite{zitzler2004indicator}, which is an indicator used in multi-objective optimization as both a proximity and diversity performance metric. Given that the optimal scenario arises when both availability and accuracy achieve a value of 1, we will use the reference point $(1,1)$ to assess performance. In this context, a higher hypervolume signifies superior model performance.

Each driving sequence, lasting up to 30 seconds and comprising numerous coordinates, was represented by the average of these coordinates to denote the geographical center of the sequence. Subsequently, the data were categorized by country and weather conditions, followed by the application of a K-means clustering algorithm to create three clusters within each group. Finally, data sharing the same label across different groups were consolidated into a singular dataset. Fig. \ref{fig:datasets} shows an example of dataset split in Sweden. The reason for this split as opposed to a conventional random split was to avoid evaluating on road segments that occur in the training set, which may be the case with a random split since the dataset contains multiple sequences of the same route. Also, the weather condition for different drives should be relatively evenly distributed in all datasets. Based on this split, dataset for training probabilistic classifier has 1343 sequences (55-hour driving); dataset for optimizing trigger function parameters has 930 sequences (38-hour driving); and the dataset for evaluating has 497 sequences (21-hour driving).

\begin{figure}
    \centering
        \includegraphics[width=\columnwidth]{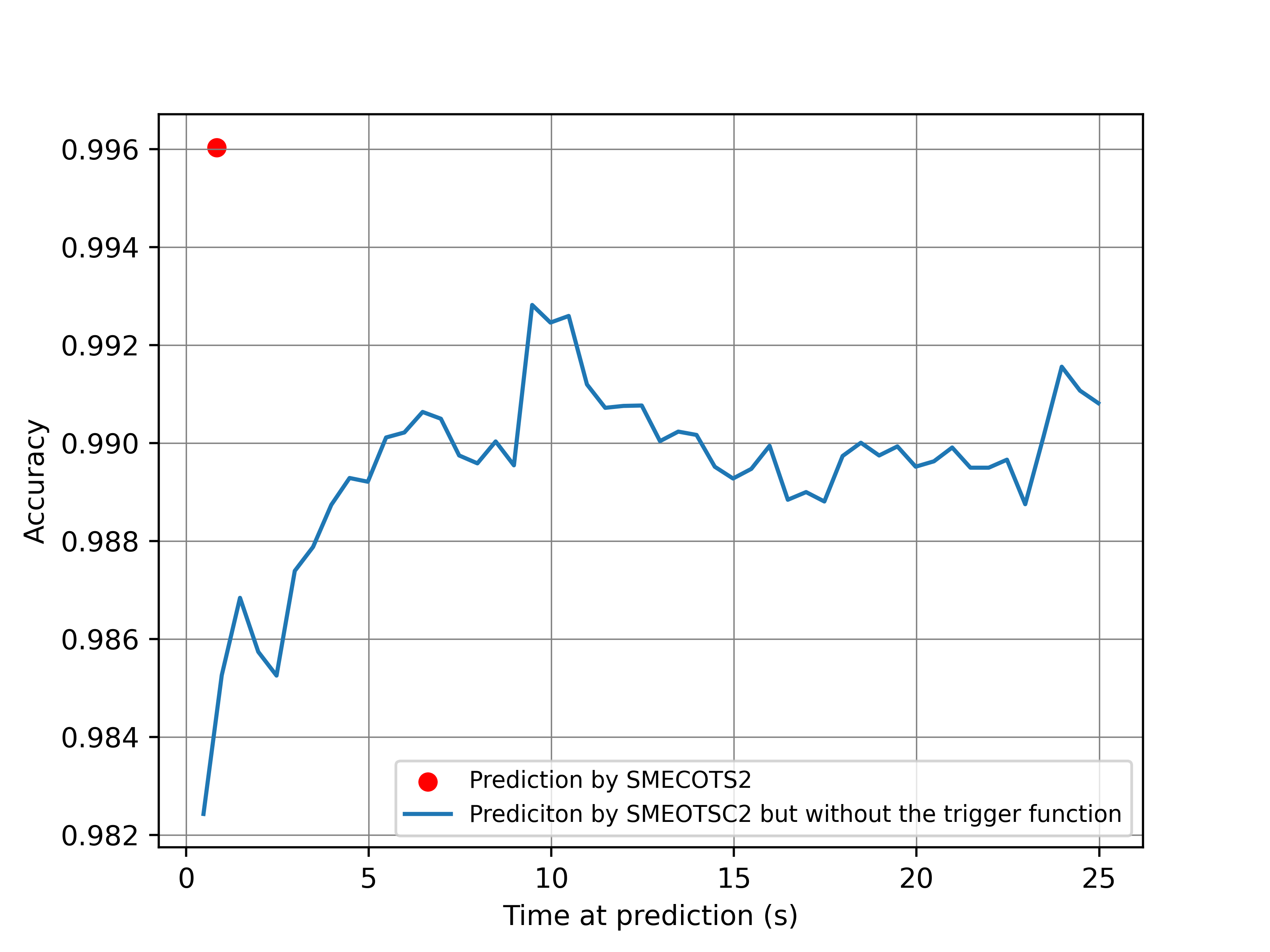}
        \caption{Compare the prediction performance of our best-performing ETSC method, SMEOTSC2, in all driving data with and without a trigger function. Without a trigger function, the method makes independent predictions at each time step. However, when incorporating the trigger function, it achieves an accuracy of 99.6\% with an average prediction time of 0.84 seconds. }
    \label{fig:time_acuracy}
\end{figure}

\begin{table}[t]
    \centering
    \caption{Performance Metrics Across Four Aspects for Each Method. The metrics, from left to right, include Earliness, Availability, Accuracy, and Hypervolume, respectively.}
    \begin{tabular}{c|c|c|c|c}
        \hline
            Methods & Ear.(s)$\downarrow$ & Ava.$\uparrow$ & Acc.$\uparrow$ & HV$\uparrow$ \\
        \hline
            METSC \cite{Mori2019EarlyTechniques} & 3.89 & 0.9286 & \textbf{0.9960} & 0.6565 \\
            MEOTSC \cite{achenchabe2023classify} & 1.53 & 0.9663 & 0.9940 & 0.7711 \\
            SMEOTSC1 & 2.44 & 0.9405 & 0.9940 & 0.7562 \\
            SMEOTSC2 & \textbf{0.84} & \textbf{0.9881} & \textbf{0.9960} & \textbf{0.8405} \\
    \end{tabular}
    \label{tab:trigger_function_comparison}
\end{table}

\subsection{Result and Analysis}
\label{sec:ra}
In Fig. \ref{fig:time_acuracy}, the significance of using the trigger function is emphasized by contrasting its performance with the accuracy obtained when predictions are made at each time step. SMEOTSC2 is developed based on \cite{Mori2019EarlyTechniques} but the trigger function is replaced with Equation~\eqref{eq:trigger4}. The red dot indicates the average prediction time and the corresponding average accuracy of that decision made by SMEOTSC2. The blue line represents the accuracy obtained by always selecting the hypothesis with the highest probability without activating the trigger function. It shows that activating trigger function not only yield higher accuracy compared to always selecting the hypothesis with the highest probability, but also facilitate earlier decisions with greater confidence.

With the dataset being used, Fig. \ref{fig:time_acuracy} points to a correlation between the duration of the prediction time and the probability of achieving an accurate decision. The analysis reveals diminishing returns associated with extended waiting times before making a prediction. While a brief delay enhances prediction accuracy, the advantages diminish as the waiting period extends. Consequently, these findings underscore the potential for striking a balance between early decision and accurate decision in lane-level localization problems.

Table \ref{tab:trigger_function_comparison} presents the four performance metrics for each method. Both SMEOTSC1 and SMEOTSC2 are developed based on \cite{Mori2019EarlyTechniques} but the trigger functions are replaced with Equation~\eqref{eq:trigger3} and Equation~\eqref{eq:trigger4} respectively.
Notably, our adapted SMEOTSC2 method demonstrates superior performance than the three other listed methods. It has the earliest prediction time (0.84 seconds), the highest availability (0.9881), and the highest accuracy (0.9960). Hypervolume (0.8405) also shows the highest results corresponding to each trigger function, serving as an evaluation metric for their performance. It's remarkable that our method can deliver predictions in under one second on average and achieve an availability of 0.9881, indicating it can accurately identify the ego-lane in 98.81\% of scenarios without compromising accuracy when compared to other methodologies.

As discussed in Section \ref{sec:lbhi}, the incorporation of a time element in the trigger functions (\ref{eq:trigger1}) and (\ref{eq:trigger2}) from METSC and MEOTSC has been identified as a limiting factor for early prediction capabilities. MEOTSC, in its adaptation to the open time series challenge, attempts to mitigate this by defining a total time step upfront and decreasing the time value as time progresses. Although this modification shows an improvement over its predecessor, it still restricts the algorithm's potential for making timely predictions. SMEOTSC1 is based on the trigger function described in Equation~\eqref{eq:trigger3}, and it retains the parameter $\gamma_3$, which is found to further hinder early-stage prediction abilities when compared to MECOTS. This suggests that maintaining a constant parameter does not enhance the algorithm's performance in the context of open time series problems, likely because a constant parameter is challenging to adapt to the time series context and struggles to accommodate temporal variations. Availability appears to be linked to earliness; SMECOTS2 not only delivers predictions in the shortest time step but also exhibits the highest availability, with MECOTS following closely in both aspects. The accuracy of the algorithms does not significantly differ, underscoring that precision in predicting the ego-line is primarily influenced by parameters $\gamma_1$, $\gamma_2$, $p'$, and $p''$, with $\gamma_3$ and the time step affecting only the earliness and availability of predictions. In our experiments, the hypervolume indicator, which measures the proximity of optimization objectives to the reference point (set at $(1,1)$), highlights that since accuracy remains consistent across the board, the algorithm boasting the highest availability (SMECOTS2) achieves the superior hypervolume.



\section{Conclusions}
\label{sec:conclusions}
In this paper, we employ Early Time Series Classification to tackle the ego-lane identification task in autonomous driving, leveraging the \textit{measurement model quality} (MMQ) metrics to represent similarities between the map and the lane markings perceived by the vehicle camera. This approach effectively balances prediction time and accuracy, ensuring prompt and confident decision-making. Our experiments in real-world traffic driving data reveal that utilizing only road lane marking geometry and lane type information, our method achieves an impressive accuracy of 99.6\% and availability of 98.81\% with an average selection time of only 0.84 seconds.

In the future, this work could be extended by incorporating additional MMQ. For instance, MMQ calculated from the position of tracked traffic signs or the position of road edges which could provide further insights into the position of the ego-lane. Similarly, tracked surrounding vehicles could offer valuable cues about the lanes they occupy, enhancing ego lane assignment algorithms.





\bibliographystyle{IEEETran}
\bibliography{Paper/references}

\end{document}